\documentclass[preprint]{aastex}
\usepackage{epstopdf}
\usepackage{graphicx}
\def\putplot#1#2#3#4#5#6#7{\begin{centering} \leavevmode
\vbox to#2{\rule{0pt}{#2}}
\includegraphics{#1}

\end{centering}}

\def\Msun{M_\odot}
\def\Lsun{L_\odot}

\slugcomment{Submitted to ApJ}

\shorttitle{M31 Red Nova}
\shortauthors{Shara et al}

\begin{document}

\title{The Red Nova-like Variable in M31 - A Blue Candidate in Quiescence}

\author{
Michael~M.~Shara,\altaffilmark{1}
David~Zurek,\altaffilmark{1}
Dina~Prialnik,\altaffilmark{2}
Ofer~Yaron,\altaffilmark{2}
and
Attay~Kovetz,\altaffilmark{2,3}
}

\altaffiltext{1}{Department of Astrophysics, American Museum of Natural
History, Central Park West and 79th street, New York, NY 10024-5192}
\altaffiltext{2}{Department of Geophysics and Planetary Sciences,
Sackler Faculty of Exact Sciences, Tel Aviv University, Ramat Aviv
69978, Israel}
\altaffiltext{3}{School of Physics and Astronomy, Sackler Faculty of
Exact Sciences, Tel Aviv University, Ramat Aviv 69978, Israel}
\altaffiltext{4}{Space Telescope Science Institute, 3700 San Martin Drive,
Baltimore MD 21218}

\begin{abstract}

M31-RV was an extraordinarily luminous ($\sim 10^6 \Lsun$)  eruptive variable, displaying very cool temperatures (roughly 1000 Kelvins) as it faded. The photometric behavior of M31-RV (and several other very red novae, (i.e. luminous eruptive red variables) has led to several models of this apparently new class of astrophysical object. One of the most detailed models is that of "mergebursts": hypothetical mergers of close binary stars. These are predicted to rival or exceed the brightest classical novae in luminosity, but to be much cooler and redder than classical novae, and to become slowly hotter and bluer as they age. This prediction suggests two stringent and definitive tests of the mergeburst hypothesis. First, there should always be a cool red remnant, and NOT a hot blue remnant at the site of such an outburst. Second, the inflated envelope of a mergeburst event should be slowly contracting, hence it must display a slowly {\it rising} effective temperature. We have located a luminous, UV-bright object within $0.4{\arcsec}$ ($1.5 \sigma$ of the astrometric position) of M31-RV in archival WFPC2 images taken 10 years after the outburst: it resembles an old nova. Twenty years after the outburst, the object remains much too hot to be a mergeburst. Its behavior remains consistent with that of theoretical nova models which erupt on a low mass white dwarf. Future Hubble UV and visible images could determine if the M31-RV analogs (in M85 and in M99) are also behaving like old novae. \end{abstract}

\keywords{binaries: eclipsing --- novae, cataclysmic variables --- binaries: close --- novae}

\section{Introduction}

One of the most puzzling stellar eruptions ever detected is due to the object known as M31-RV. This variable, in the nuclear bulge of M31, \citep{ric89} brightened in mid-1988 to almost $10^6 \Lsun$. At the peak of its outburst it rivaled the most luminous stars in the Local Group, and was as bright as the brightest classical nova ever seen. The 2-3 month interval that M31-RV spent brighter than Mbol$= -6$ is consistent with it being a very luminous nova. The old stellar population at the site of M31-RV is also consistent with the classical nova hypothesis \citep{bon06}. However, as M31-RV faded, its spectrum evolved from that of an M0 supergiant to M5 and then to late M \citep{ric89, mou90}. Fading classical novae are not expected to display M supergiant colors or spectra. On the contrary, after ejecting of order $10^{-5}\Msun$ via a thermonuclear runaway on their surface, many models of classical nova white dwarfs remain hotter than $10^{5}$ K \citep{yar05,pri79} for decades. Thus post-novae are expected to be hot and blue. In Figure 1 we show the time-dependent luminosities and temperatures of the erupting white dwarf in a nova model.

Some old novae are indeed observed to remain very hot for years, while others eject their envelopes quickly and cool on a timescale of a year or less, as shown by \citet{van01} and \citet{ori06}. The values of all of the reddening-corrected, optical and near-infrared colors of decades-old novae tend to cluster around 0 \citep{szk94}. This is because the optical emission from old novae is usually dominated by the binary system's accretion disk, displaying an effective temperature of about 10 kKelvins. The spectra of M31-RV are representative of much cooler material (1000 - 3000 Kelvins). If M31-RV wasn't a nova, then what could it have been?

An even more luminous, but similarly red variable in the Virgo cluster galaxy M85 \citep{kul07} was seen in 2006. Like M31-RV, the M85 optical transient (M85- OT) was not associated with any bright star forming region \citep{ofe08}. The g- and z-band absolute magnitudes of the progenitor were fainter than about -4 and -6 mag, respectively, corresponding to an upper limit for a progenitor (main sequence) mass of $7 M_\sun$. Another remarkably luminous "red nova" has recently been detected in the Virgo spiral galaxy M99 \citep{kmk10}.

V838 Mon is the best-studied Galactic eruptive variable which rivals M31-RV and M85-OT in outburst peak luminosity; it was also very cool and red as it faded; many details are reported in \citet{cor07}. This object resembled M31-RV both in luminosity and spectral evolution, cooling to an L-type supergiant \citep{lyn04} and displaying both a dust shell and an extended, evolving light echo \citep{mun02}. However, the stellar environments of M31-RV and M85-OT stand in sharp contrast to that of V838 Mon \citep{bon03}. V838 Mon has been associated with a group of B stars, and is unresolved by the Hubble Space Telescope at the $0.2\arcsec$ level from a B3 dwarf, though the progenitor of the eruptive variable was not luminous \citep{afs07}. Since V838 Mon seems to be coeval with the B stars it must be far too young to have evolved into a nova binary system. The deep eclipses of the B3V star noted by \citet{mun07}, \citet{gor08}, and \citet{kol09} are suggestive of accretion and interaction in a wide binary.

A detailed search for historical analogs of M31-RV has been carried out by \citet{kim07}. Among the candidates are the Galactic eruptive variable V4332 Sgr, which has displayed very cool spectra and suggestions of a thick disk obscuring a central cool star \citep{kam10}; unfortunately its distance and luminosity are highly uncertain \citep{mar99}. A new, possibly related object (V1309 Sco) has recently been proposed \citep{mas10}.

Do V838 Mon and other very red, luminous eruptive variables represent a new and rare class of astrophysical object? The photometric and spectroscopic behaviors of the red nova outbursts are very similar, as noted by \citet{mnv07} and \citet{bos04}. The frequently claimed inability of classical novae to produce cool, red spectra, and the young environment of V838 Mon, have led to a plethora of alternative models. These include the planet-capture model \citep{ret03} and \citep{ret06}; born-again objects \citep{law05}; a thermonuclear shell flash in the outer layers of a highly evolved, young massive star \citep{mun05}; a low-mass AGB star experiencing thermal pulses going into the post-AGB phase \citep{van04}; a mass-transfer episode from an extreme asymptotic giant branch star to a main sequence companion \citep{kmk10}; and the electron-capture supernova model \citep{bot09}.

Most of these models have been extensively reviewed and critiqued by \citet{sok06}, who maintain that all but one fail to explain the extensive observations of V838 Mon and M31-RV. That one notable exception is the mergeburst model \citep{sok06}, which describes the merger of two stars. A significant fraction of the energy released during such a merger must inevitably swell the merger product envelope, cooling it to 2000 K or less. To quote Soker and Tylenda (2006): ``Violent and luminous mergers, which we term mergebursts, can be observed as V838 Monocerotis-type events, where a star undergoes a fast brightening lasting days to months, with a peak luminosity of up to $10^{6} \Lsun$ followed by a slow decline at very low effective temperatures". These clean, simple predictions are testable, and they sharply differentiate the mergeburst model from models of classical novae.

A white dwarf which powers a classical nova is inevitably left with a thin, very hot shell of hydrogen after mass ejection ends. Thus the central stars of classical novae initially display high effective temperatures and appear very blue after their eruptions subside and their ejecta become optically thin. As the remnant hydrogen-rich shell on its white dwarf cools, the effective temperature displayed by an old nova must decrease, and the system must become much redder on a timescale of decades. The two key mergeburst observables behave in opposite fashion: a mergeburst's bloated envelope must initially produce a cool, red remnant, and that remnant must get hotter (on a timescale of millenia) as the swollen envelope contracts \citep{sok06}. This sharp contrast in behavior suggests that recovery of M31-RV,  M85 OT and similar objects years and decades after their eruptions can be a crucial test of the mergeburst models.

In the next section we note the satellite and ground-based images that we have examined in a search for the remnant of M31-RV, and show evidence for a very luminous, variable blue star at its position. We then present a new series of HST images of this star, showing both its current blue color and continued fading. Just as important, the longer baseline we now have shows that our M31-RV candidate is unquestionably cooling. Both of these observations suggest that M31-RV and V838 Mon may be very different phenomena. Both of these observations are in disagreement with the predictions of the mergeburst scenario, and all the other scenarios noted above, except for extreme nova models. We compare the observations of M31-RV to the theoretical predictions of new classical nova models, and find remarkably good agreement. We conclude by outlining future observations to further test these models' predictions. An accompanying paper \citep{sha10} details new nova models (with very massive WD envelopes) which mimic, with considerable fidelity, the ``red nova" M31-RV.

\section{M31-RV's position}

We have searched ground-based telescopes' archives for images of M31-RV in quiescence. The very best available are those from the Canda-France-Hawaii Telescope (CFHT), with seeing in the $0.5\arcsec$ range. Even these images are hopelessly confused by the severe crowding in M31. We have also searched the Galaxy Explorer (GALEX) Far-Ultraviolet (FUV) and Near-Ultraviolet (NUV) images of M31. While there are a few FUV pixels significantly above background at the position of M31-RV, the $6\arcsec$ pixels of this satellite are far too coarse to yield a reliable detection. There are no X-ray sources close to the position of M31-RV. We conclude that only with the Hubble Space Telescope (HST) is there any realistic chance of resolving and recovering the remnant of M31-RV.

A position for M31-RV (with a one $\sigma$ error of $0.27\arcsec$) has been determined by \citet{bon06} from archival CCD frames of that star in eruption. This was done with three Kitt Peak National Observatory (KPNO) CCD frames, showing M31-RV in eruption. These frames were used to determine a position for the variable, accurate to $0.04\arcsec$ in each coordinate, relative to nine bright astrometric stars in the NOMAD catalog \citep{zac04}. The same nine stars were located on deeper KPNO 4 meter telescope frames, which were then matched to several epochs of HST images of the same field in M31.

Because the results of this investigation depend critically on the accuracy of the astrometric position that we can determine for M31-RV, DZ has independently re-measured the position of M31-RV. The same CCD frames of M31-RV in eruption (taken by R. Ciardullo) were analyzed in similar fashion as above, except that different astrometric standard stars were located on deep CFHT archival images, which were in turn matched to HST images of the same field in M31. The position determined by DZ is 00:43:02.438 and +41:12:56.24  (J2000). The M31-RV position determined by \citet{bon06} is 00:43:02.433 and +41:12:56.17 (J2000). This excellent agreement (to within $0.09\arcsec$), using independently selected astrometric standard stars and different archival 4 meter telescope images, strongly supports both the $0.27\arcsec$ error and the position of M31-RV claimed by \citet{bon06}, both of which we adopt.

\section{The Archival HST Data and Blue Candidate}

Table 1 details all archival HST images (up to 2008) covering the field of M31-RV. Figures 2a, 2b, 2c and 2d are HST archival images of the neighborhood of M31-RV. Figures 2a and 2b are from 2003 and 2004 through the blue filter F435W with the Advanced Camera for Surveys (ACS), while Figures 2c and 2d were taken in 1995 through the two UV filters F300W and F170W (with the Wide Field and Planetary Camera (WFPC2)), respectively. The extraordinary crowding in M31 is apparent, especially in Figures 2a and 2b. The archival F300W and F170W images are much shallower, but enough objects are seen in common (and circled in both images) that there is no doubt that we are looking at the identical field in each filter, and that this field corresponds to the position of M31-RV determined in the previous section.

Figures 3a, 3b, 3c, and 3d are archival HST images taken at different epochs and with different filters, of the part of Figure 2 that is indicated with a square. The astrometric position of M31-RV \citep{bon06} is indicated with an ``x", with one and $3 \sigma$ error circles surrounding that position. It is clear in Figure 3a that a very bright star was present within $1.5 \sigma$ ($0.41\arcsec$) of that astrometric position for M31-RV in 1994, about 7 years after the eruption. The star appeared in the HST F300W (U band) images at m(F300W) $=21.4$. This corresponds to an unreddened absolute magnitude M(F300W)$=-2.6$,  or 1000 $\Lsun$ (in excellent agreement with the luminosity predicted in Figure 1 for a nova 7 years after outburst). The observed reddenings of globular clusters near the nucleus of M31 (next paragraph) suggest a dereddened Mo(F300W) that is at least a magnitude more luminous. Unfortunately, these F300W images have never been repeated, and the same field was not imaged in other filters in 1999. The rarity of such luminous and hot, UV-bright objects in M31 is immediately apparent from Figure 3a. A main sequence dwarf star would have to be of early type B (with a mass of at least 8 M$_\sun$) to be this luminous, and display U-B ($\sim -0.3$), B ($\sim 21.7$), and an effective temperature of at least 20 kKelvins.

Five years later, in 1999, the same object appeared at a m(F555W)$=23.33$ and m(F814W)$=23.28$, corresponding to V-I$=0.05$. Most globular clusters close to the nucleus of M31 display $0.3 < E(V-I) < 0.5$ \citep{bar00}, so that (V-I)o for the blue star we are considering is in the range $-0.25 < (V-I)o < -0.45 $. This immediately implies a photospheric temperature T $> 40$ kKelvins, far too blue and hot for any mergeburst model (with expected apparent temperatures of $1000$ to $3000$ K.) Even ignoring reddening, the observed color in 1999 (about 10 kKelvins) of our candidate for M31-RV rules it out as a mergeburst. The apparent luminosity ($\sim 250\Lsun$) of the object in 1999 is again in good agreement with that expected from a nova 11 years after eruption, as seen in Figure 1.

Finally, the object was seen at m(F435W)$=24.1$ in both 2003 and 2004, so (at least during that one year-long interval) it was nearly constant in brightness at about $100\Lsun$. This is again in reasonable agreement with Figure 1, which also shows that an old nova is expected to fade only very slowly once it is $15-16$ years past eruption.

The object must be variable. If not, then we could derive its color from the 1994 (F300W) and 2004 (F435W) magnitudes. The resulting F300W - F435W color is $-2.7$, which is unphysical. It is possible that the unphysical color noted above  is due to strong emission lines in the ejecta spectrum. Clearly, it is important to image this object simultaneously in multiple passbands to determine its present colors, effective temperature and luminosity. If our candidate is M31-RV and it was a mergeburst then it must slowly get hotter as its bloated envelope contracts.  Conversely, if our candidate for M31-RV was a classical nova, then it must continue to fade and to cool on a time-scale of decades in accordance with Figure 1.

Figure 4 is our F555W versus F555W-F814W color-magnitude diagram of the field of M31-RV, taken in 1999. This is the only HST archival epoch where images in two colors are simultaneously available. Our candidate stands out as the brightest (and by far the UV-brightest) object in its vicinity.

Of course it is always possible that our blue candidate might have nothing to do with M31-RV. We can rule out a background supernova as it would have faded away completely between 1994 and 1999. We can rule out an M31 RR Lyrae star as the object seen in 1994 is much too luminous to be one. We can also rule out the planet-capture model, born-again objects, a thermonuclear shell flash in the outer layers of a highly evolved, young massive star, a low-mass AGB star experiencing thermal pulses going into the post-AGB phase and a mass-transfer episode from an extreme asymptotic giant branch star to a main sequence companion; all of these models leave very cool remnants that are far redder than the blue object we have descrived at the site of M31-RV.

We cannot rule out an erupting intergalactic dwarf nova that lies between the Milky Way and M31, within $0.41\arcsec$ of M31-RV. Barring this unlikely coincidence we conclude, on the basis of color and brightness, that M31-RV could well be an old nova, and that it cannot be a mergeburst, or any of the other models listed above. The hypothesis that M31-RV is the prototype of a new class of astrophysical phenomenon - mergebursts - is refuted by the 1999 color of the observed remnant if our candidate is, in fact, M31-RV.

\section{New HST Observations}
To further constrain the nature of M31-RV we requested, and were granted, 5 orbits of HST time to re-observe M31-RV on 26 July 2008. A log of our new observations is given in Table 1. The aging WFPC2 camera of HST did not enable us to image quite as deeply as at previous epochs, but the new data are still extremely useful in characterizing M31-RV two decades after it erupted. A mosaic of the F300W, F439W, F555W and F814W images is shown in Figure 5.

Comparing Figure 3a with Figure 5a we see, in the F300W images, that the UV-bright object seen $0.41\arcsec$ from the nominal position of M31-RV in 1995 has faded almost to the WFPC2 detection limit in 2008. The fading object is at least 2.8 magnitudes (a factor of 13.2) fainter in 2008 compared with 1995. This rules out a luminous field star that is not highly variable. It supports the suggestion that this variable is really M31-RV. The object became considerably redder, too; it displayed V-I = 0.05 in 1999 and V-I = 0.44 in 2008. The importance of this color change cannot be overstated. A mergeburst must slowly become {\it hotter} as its bloated envelope contracts. M31-RV is observed to have become much {\it cooler}.

The corresponding dereddened values for M31-RV are $-0.45 < (V-I)_0,1999 < -0.25 $ and $-0.06 < (V-I)_0,2008 < 0.14 $.  The 2008 (B-V) color is 0.47, corresponding to  $0.07 < (B-V)_0,2008 < 0.35$ . The latter dereddened (2008) values of $(V-I)_0$ and $(B-V)_0$ correspond to an early A star, with an effective temperature $\sim 8000 $ Kelvins, still far too hot for a mergeburst. The largest published compilation of old nova colors is that of \citet{szk94}. She showed that the dereddened values of (B-V), (V-R) and (V-J) all cluster around zero, corresponding to an effective temperature of  $\sim 10 $ kKelvins, in agreement with what we observe for our candidate. A very red mergeburst would display colors with values $> 2$ .

While our candidate has faded dramatically in the F300W filter, it has remained essentially constant in the F435W, F555W and F814W filters (see Table 1 and Figure 6).

\section{A Second Look at Classical Novae}

The observations reported here of M31-RV's observed colors are very problematic for all of the models except the nova model listed in the Introduction. The observation that M31-RV's Galactic counterpart V838 Mon appears to be a member of a young group of B stars is very problematic for a classical nova model for V838 Mon. It is not inconceivable that there are two different phenomena at work - a mergeburst in the case of V838 Mon and an extreme classical nova in the  case of M31-RV.  Thus carrying out a reexamination of the nova model is in order to see if the extremely red colors and high luminosity of M31-RV can be produced by a classical nova.

The outburst characteristics of a nova (peak brightness, ejected mass and velocity, color and temperature) are determined by the underlying white dwarf mass, temperature, and accretion rate. The most extensive set of nova simulations covering these three parameter values is due to Yaron et al (2005). Both they and \citet{ibe94} noted that a little-studied corner of nova phase space (low white dwarf mass, low accretion rate, cold white dwarf) leads to unusually massive hydrogen envelopes ($\sim10^{-3}\Msun$) before an eruption occurs. As they slowly expand, these massive shells remain optically thick to radii of order 10 times larger than those of most other novae (with shells of order 100 times more massive). This can account qualitatively for the very cool, red spectra of objects like M31-RV without invoking new astrophysical phenomena.

These new nova models do not yet include opacities appropriate for the very low temperatures (1-5 kKelvins) observed in M31-RV, and thus they produce peak luminosities that are still a factor of two less than observed in M31-RV. At the low temperatures observed in M31 RV's ejecta, recombination decreases the number of free electrons dramatically, with a consequent decrease in the ejecta opacity. This decrease in opacity would permit the rapid ``leakage" of photons out of the expanding, cool envelope that would give rise to luminosity spikes of up to $10^7 \Lsun$ \citep{sha10}

\subsection{Summary and Conclusions}

We have shown that, in 1994, a very luminous ($\sim 1000 \Lsun $), UV-bright candidate existed very close to the site of the 1988 eruption of M31-RV. We have also been able to follow up with sufficient sensitivity with HST to show that the same blue object was at least 13 times fainter in F300W in 2008 than it was in 1994. Over the past 20 years the object has not only faded, but it has also become much redder as predicted and observed for old novae.  The extremely blue colors of M31-RV (at least 40 kKelvins in 1999 and about 8 kKelvins in 2008) are incompatible with all the models proposed for V838 Mon (including models of mergebursts) , but in agreement with models and observations of post-novae.

\acknowledgments

\pagebreak

\begin{deluxetable}{cccccc}
\tablecolumns{6}
\tablewidth{0pc}
\tablecaption{HST archival Data}
\tablehead{
\colhead{Instrument} & \colhead{Filter} & \colhead{Exposure Time (sec) } & \colhead{Date (mm/dd/yyyy)} & \colhead{MJD} & \colhead{Magnitude}}
\startdata
WFPC2 & F814W & 10400.0 & 07/24/1999 & 51382.81612849 & $23.28 \pm 0.03$ \\
WFPC2 & F814W & 2400.0 & 07/26/2008 & 54673.70436972 & $22.95 \pm 0.05$ \\
WFPC2 & F555W & 7200.0 & 07/23/1999 & 51382.55293404 & $23.34 \pm 0.03$ \\
WFPC2 & F555W & 800.0 & 07/26/2008 & 54673.71409194 & $23.39 \pm 0.06$ \\
ACS/WFC & F435W & 2200.0 & 12/25/2003 & 52998.89373337& $24.47 \pm0.02$ \\
ACS/WFC & F435W & 2200.0 & 10/02/2004 & 53280.17994121 & $24.25 \pm0.02$ \\
WFPC2 & F439W & 1600.0 & 07/26/2008 & 54673.72034194 & $23.86 \pm0.1$ \\
WFPC2 & F300W & 2600.0 & 12/05/1995 & 50056.18491969 & $21.40 \pm 0.1$ \\
WFPC2 & F300W & 2400.0 & 07/26/2008 & 54673.72728639 & $<24.2$ \\
WFPC2 & F170W & 10800.0 & 12/05/1995 & 50056.25366969 & Not Detected \\
ACS/SBC & F140LP & 2552.0 & 07/26/2008 & 54674.43248819 & Not Detected \\
\enddata
\end{deluxetable}

\pagebreak

\begin{figure}[t]
\figurenum{1}
\epsscale{0.8}
\plotone{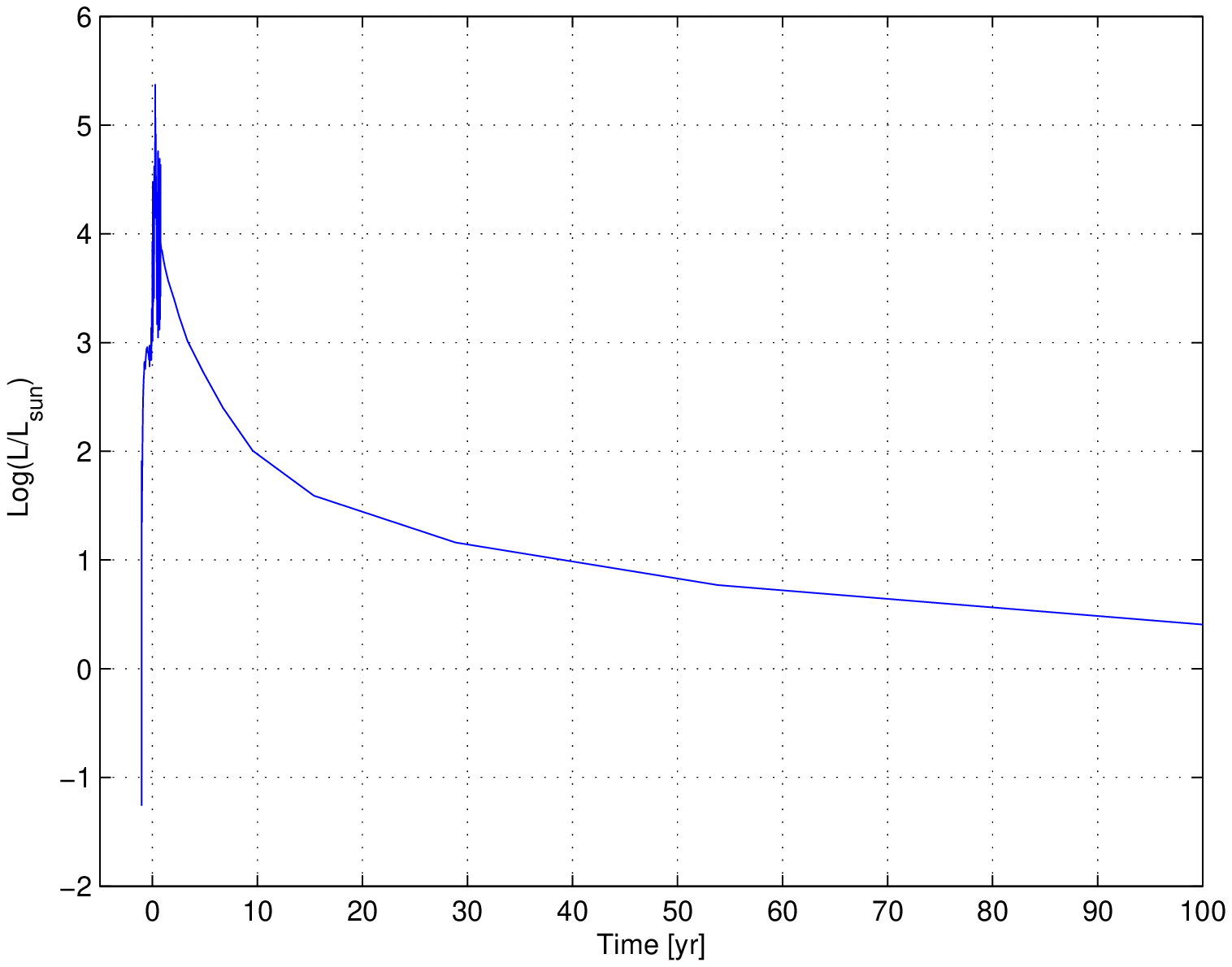}
\plotone{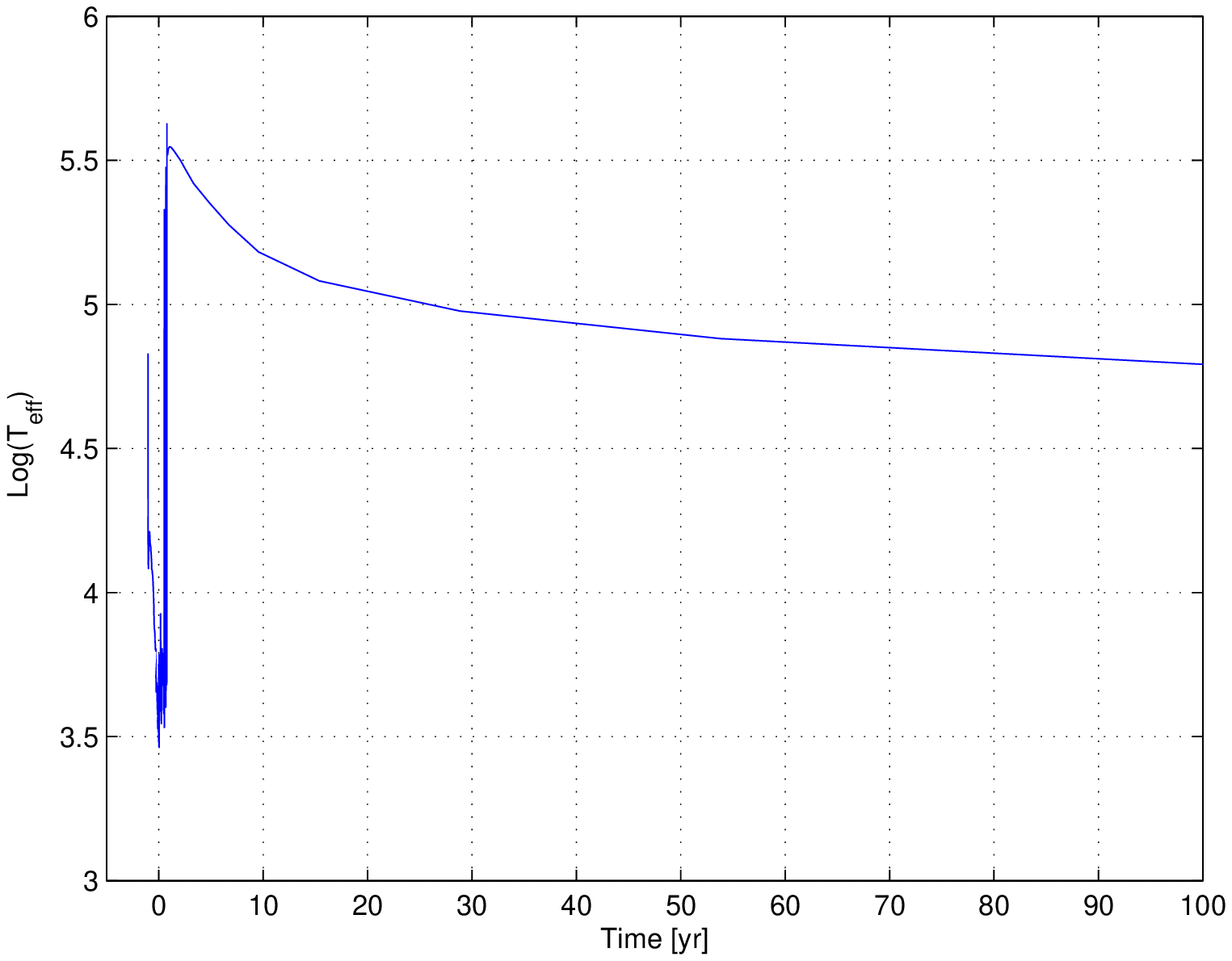}
\caption{Top: The predicted luminosity of a classical nova model during the century after eruption. The underlying white dwarf of $0.5 \Msun $ accretes hydrogen from its companion at the rate of $10^{-11} \Msun/yr$. Bottom: The predicted effective temperature of the white dwarf during the century after a nova eruption. A decade or longer after eruption, nova systems are observed to be much cooler than shown above (displaying effective temperatures of about 10 kKelvins) because their light output is dominated by their relatively cool accretion disks. }
\end{figure}


\begin{figure}[t]
\figurenum{2a}
\epsscale{0.8}
\plotone{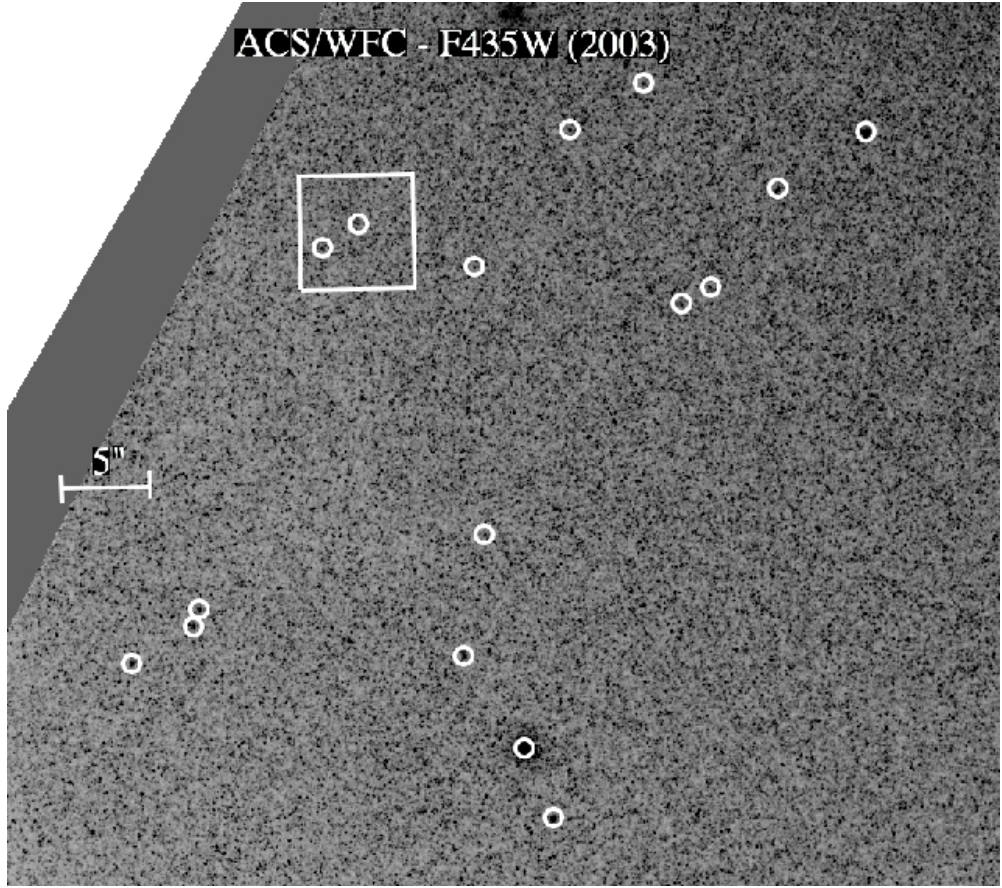}
\caption{(a) This image is taken with the F435W filter and HST/ACS/WFC in 2003. The circled objects are sources which are easily identified in the images 2a and 2b. The square region is the area in the close-up images in figure 3.  (b) This image is taken with the F435W filter and HST/ACS/WFC in 2004. These sources allow us to pinpoint the position of our candidate Red Variable in the F300W which is the circled object in the center of the square. The scale bar at the top left is $5\arcsec$. (c) This image is taken with the F300W filter and HST/WFPC2 in 1994. The circled objects are sources which are easily identified in both images 2a and 2b. The scale bar at the top left is $5\arcsec$. (d) This image is taken with the F170W filter and HST/WFPC2 in 1994. }
\end{figure}

\begin{figure}[t]
\figurenum{2b}
\epsscale{0.8}
\plotone{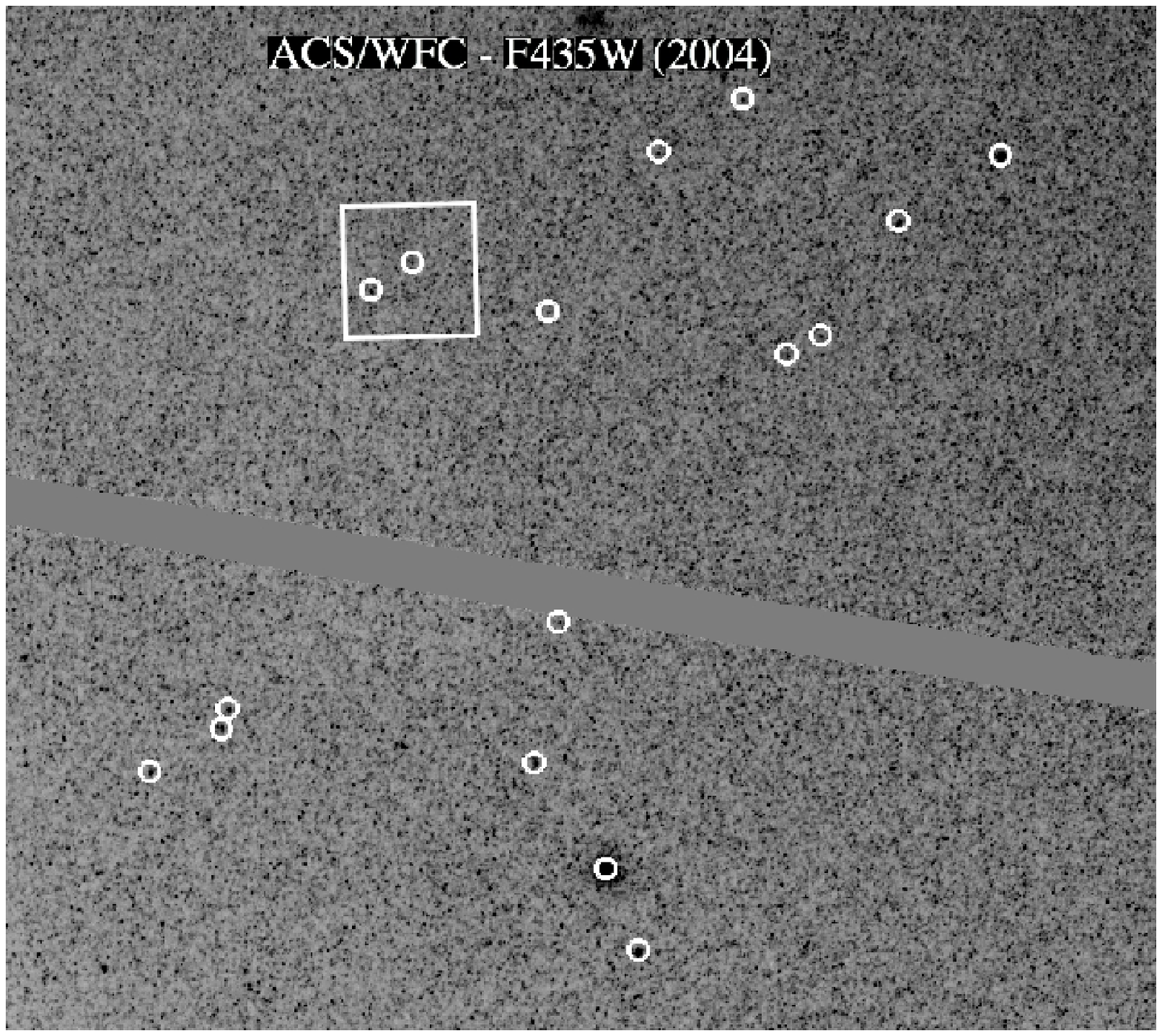}
\caption{continued}
\end{figure}

\begin{figure}[t]
\figurenum{2c}
\epsscale{0.8}
\plotone{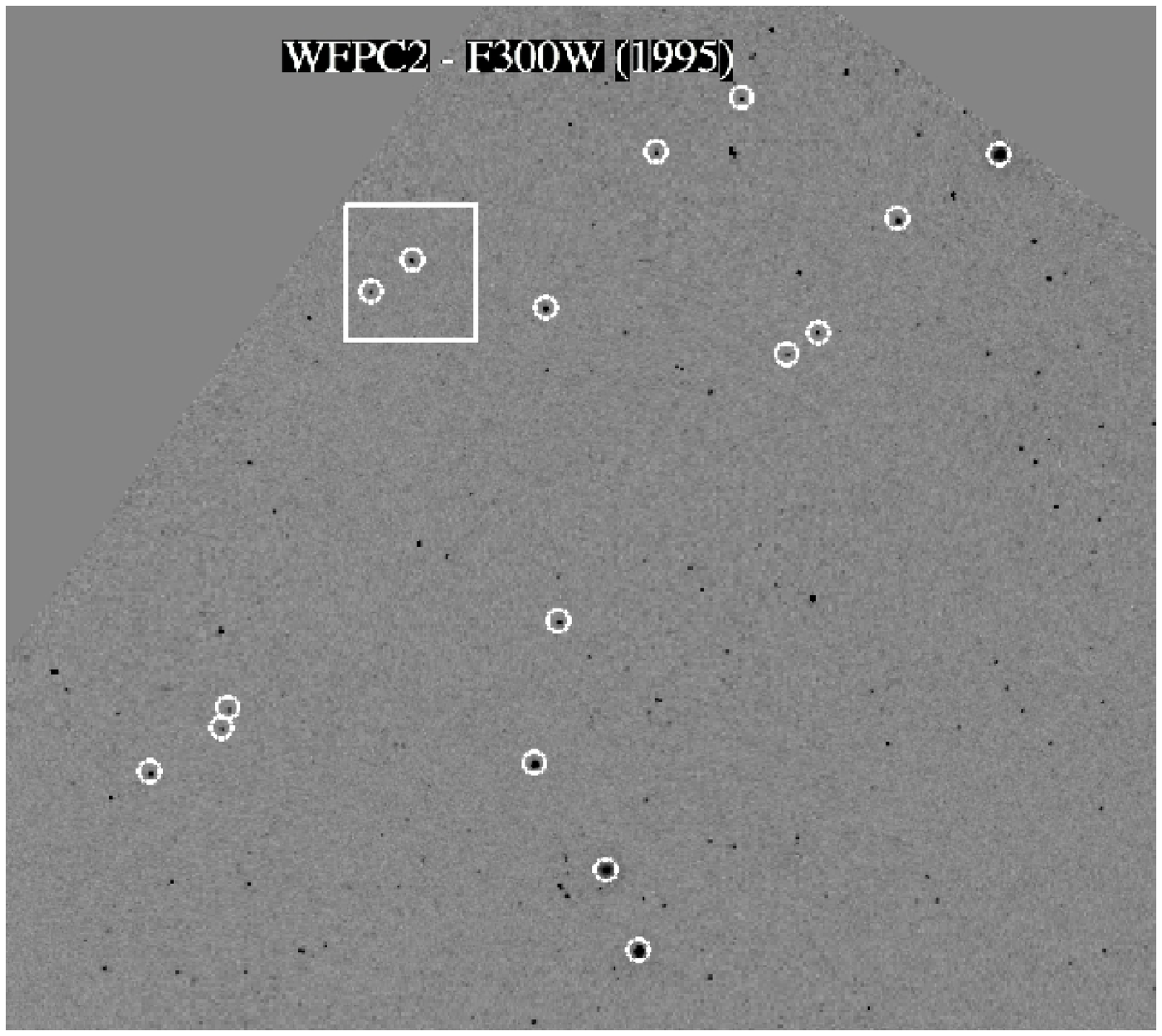}
\caption{continued}
\end{figure}

\begin{figure}[t]
\figurenum{2d}
\epsscale{0.8}
\plotone{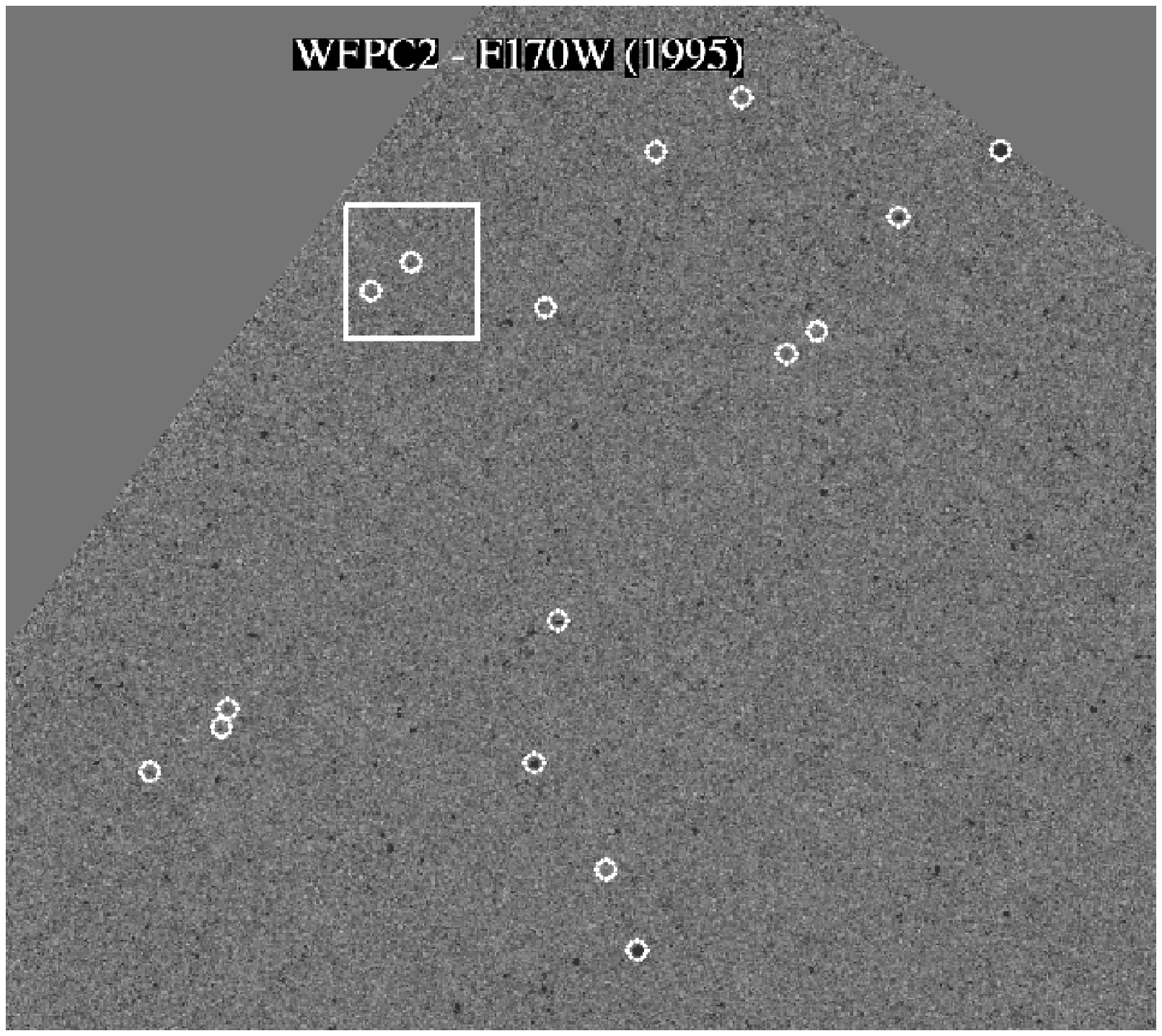}
\caption{continued}
\end{figure}

\begin{figure}[h]
\figurenum{3}
\begin{center}
{\includegraphics[scale=.39]{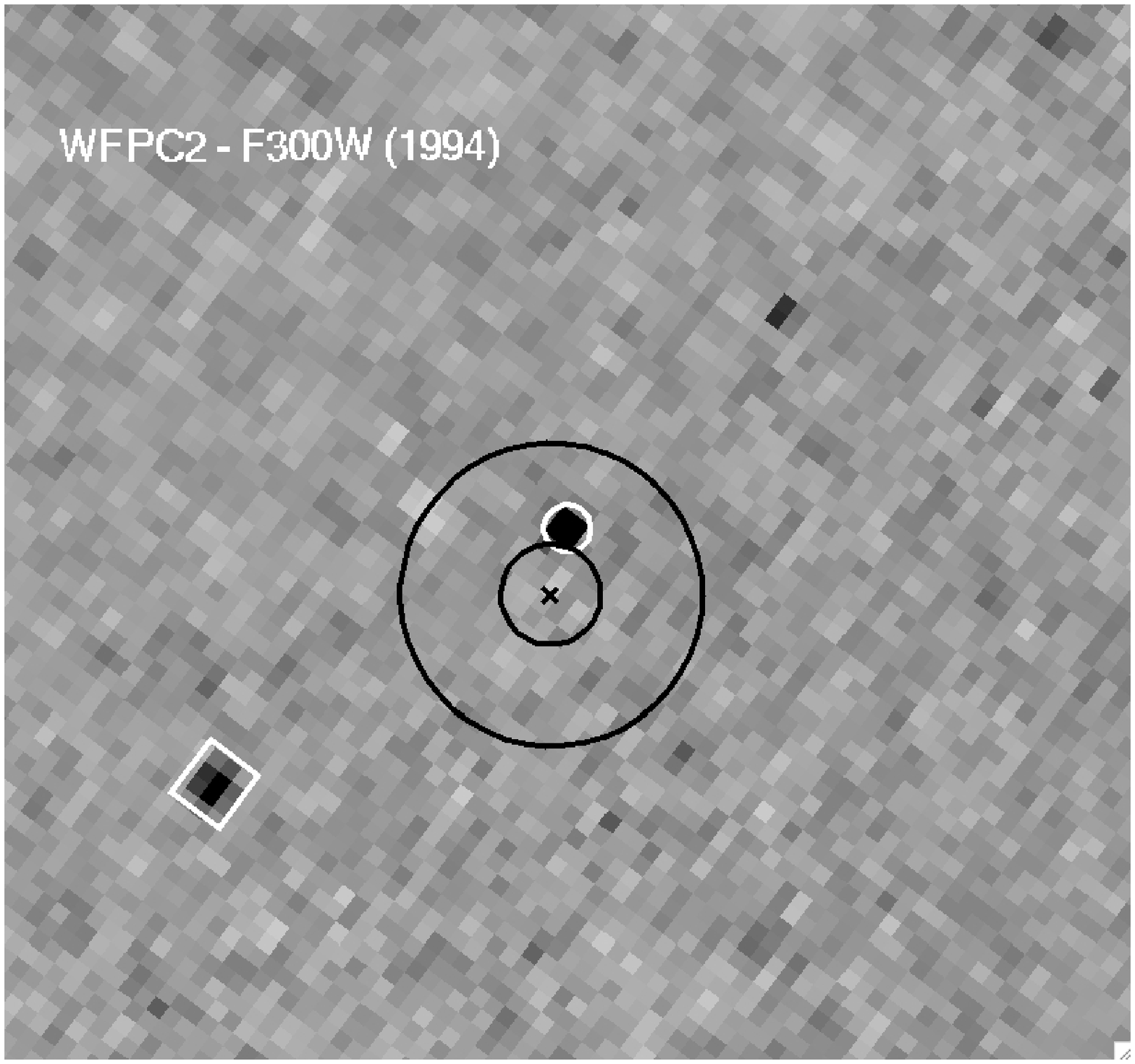}
 \includegraphics[scale=.39]{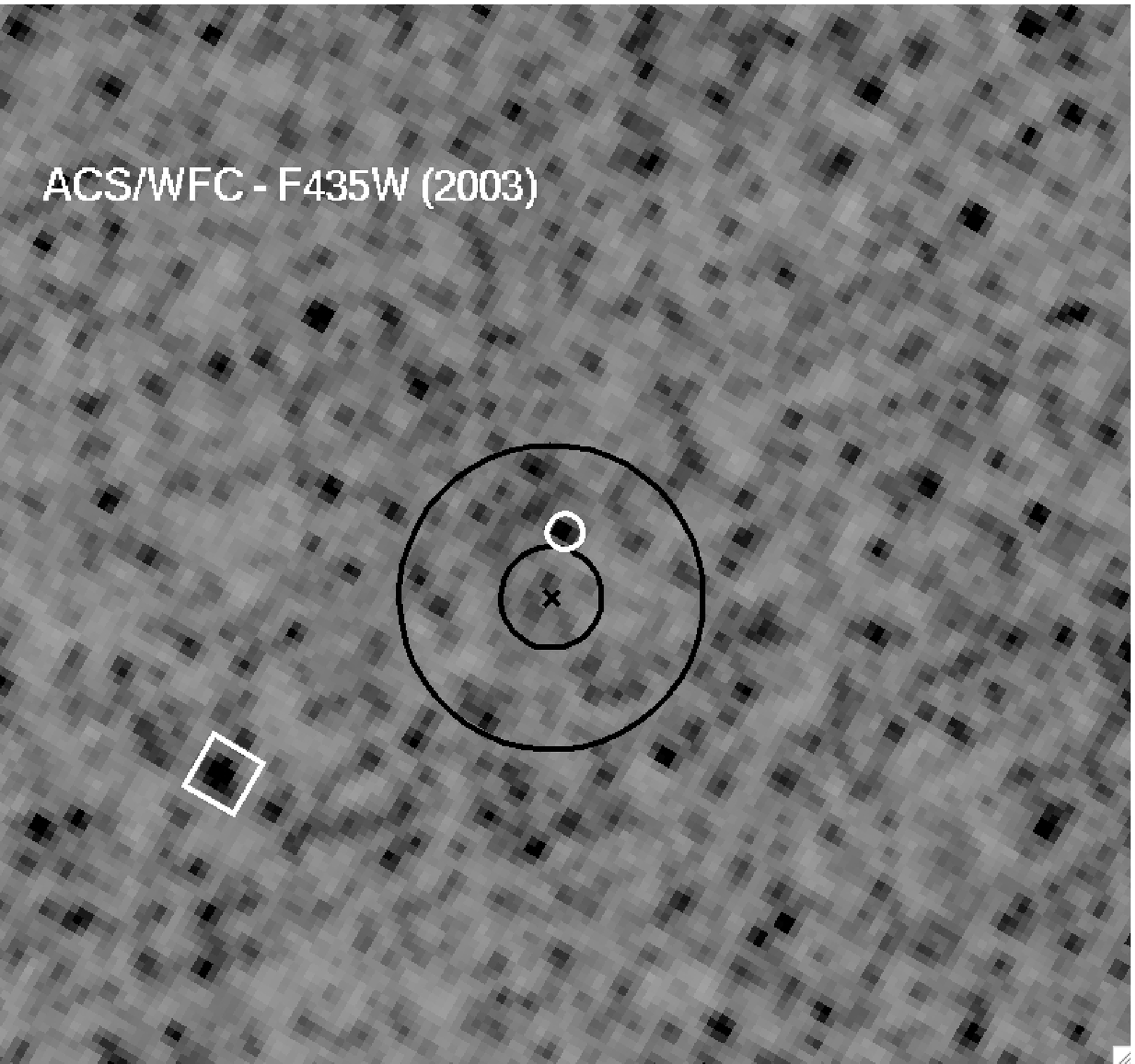}}
{\includegraphics[scale=.39]{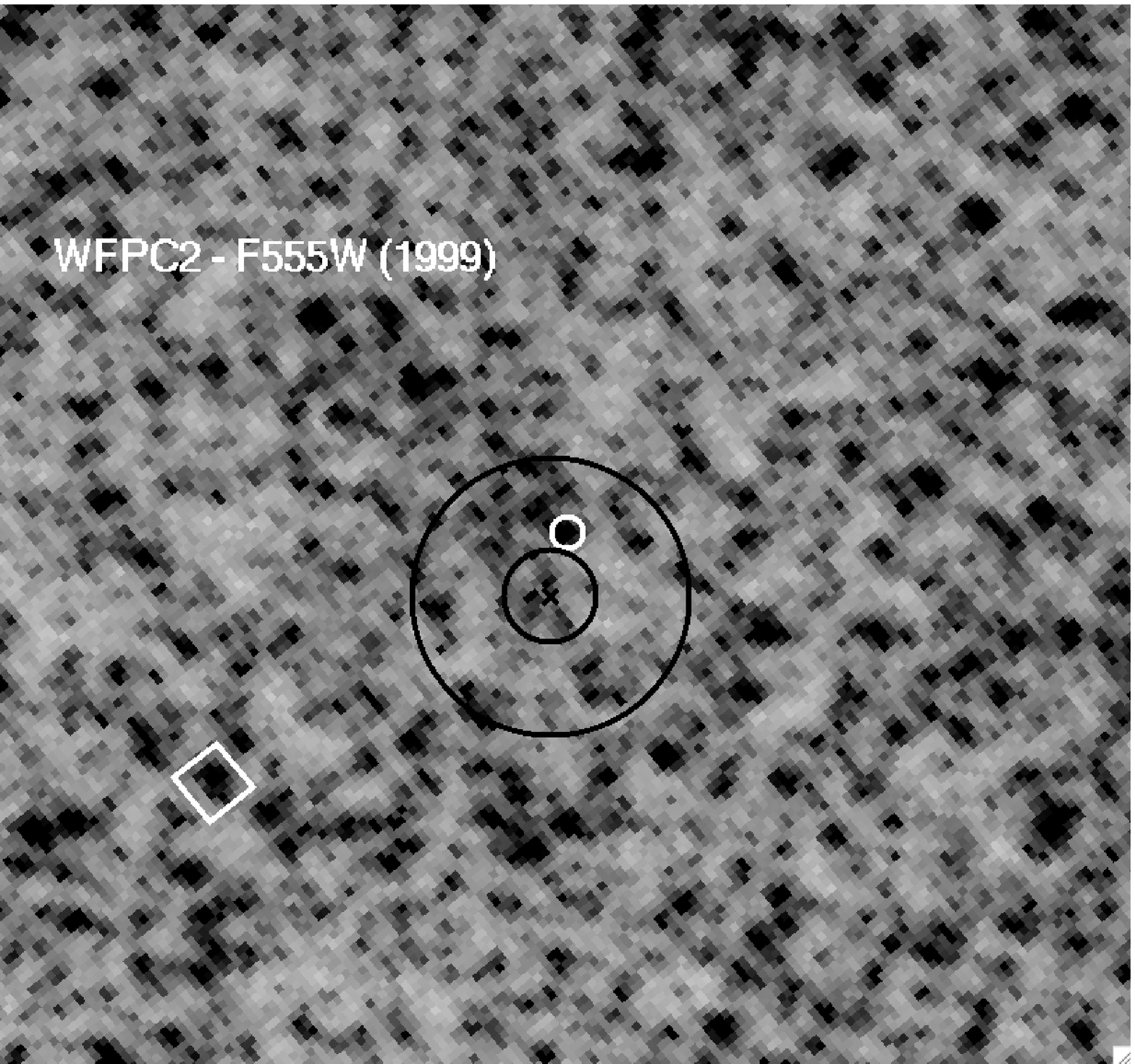}
 \includegraphics[scale=.39]{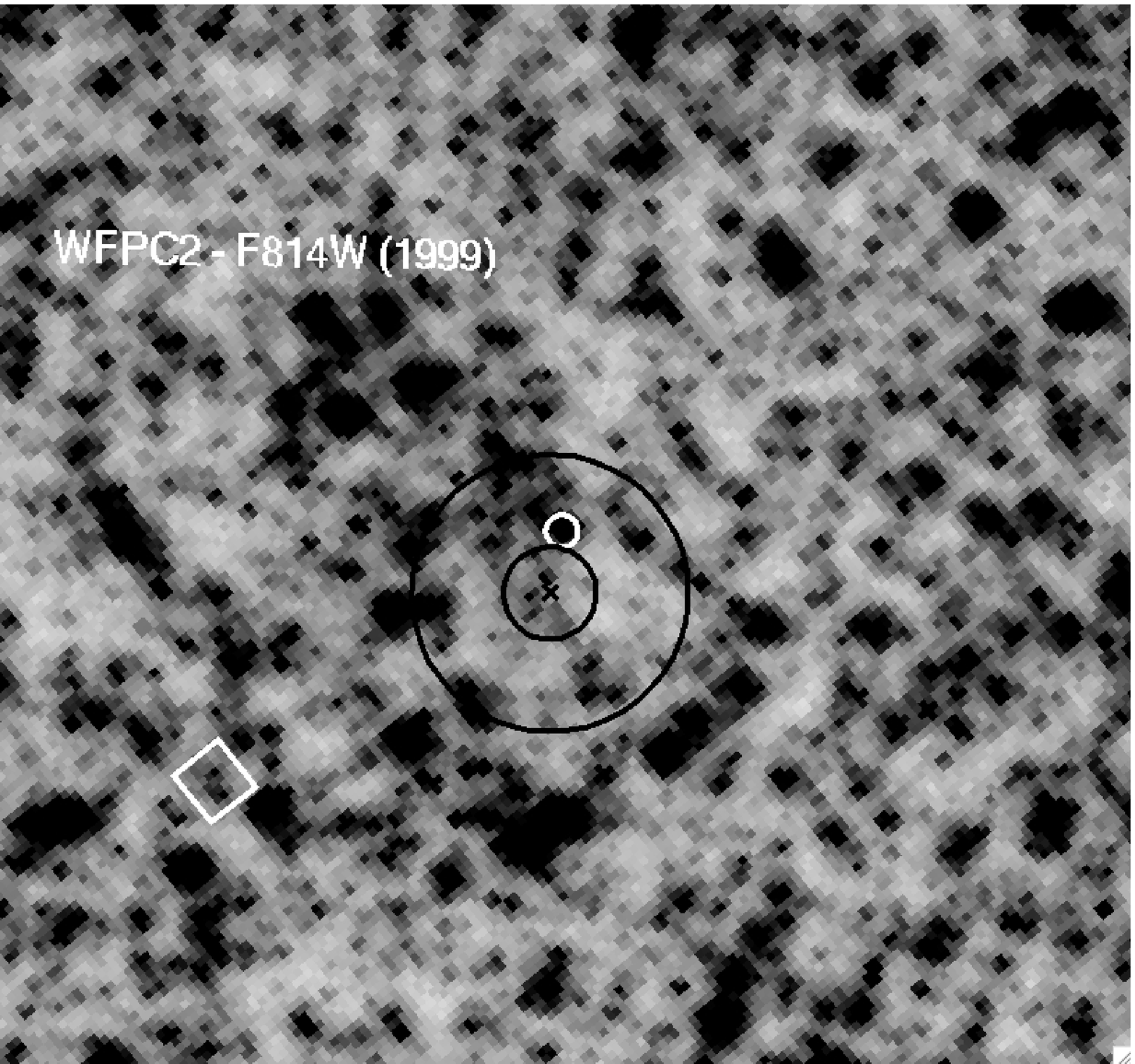}}
\end{center}
\caption{A mosaic of the images F300W (WFPC2 - 1994), F435W (ACS/WFC - 2003), F555W and F814W (WFPC2 - 1999). The x in each of the images is the position determined by Bond \& Siegel (2006) for the ``red nova'' in M31. The inner and outer circles are the 1 and 3 $\sigma$, respectively, positional accuracy of the position of Bond \& Siegel (2006). The inner circle has a radius of $0.27\arcsec$ and the outer circle has a radius of $0.81\arcsec$. The white circle surrounds the blue source described in the text. A white square encompasses the nearest fiducial star SE of the Bond \& Siegel (2006) position. Several other fiducial stars (outside the FOV shown here but seen in figure 2) assure that we have correctly identified the field of the sparsely populated F300W image.}
\end{figure}




\begin{figure}[t]
\figurenum{4}
\plotone{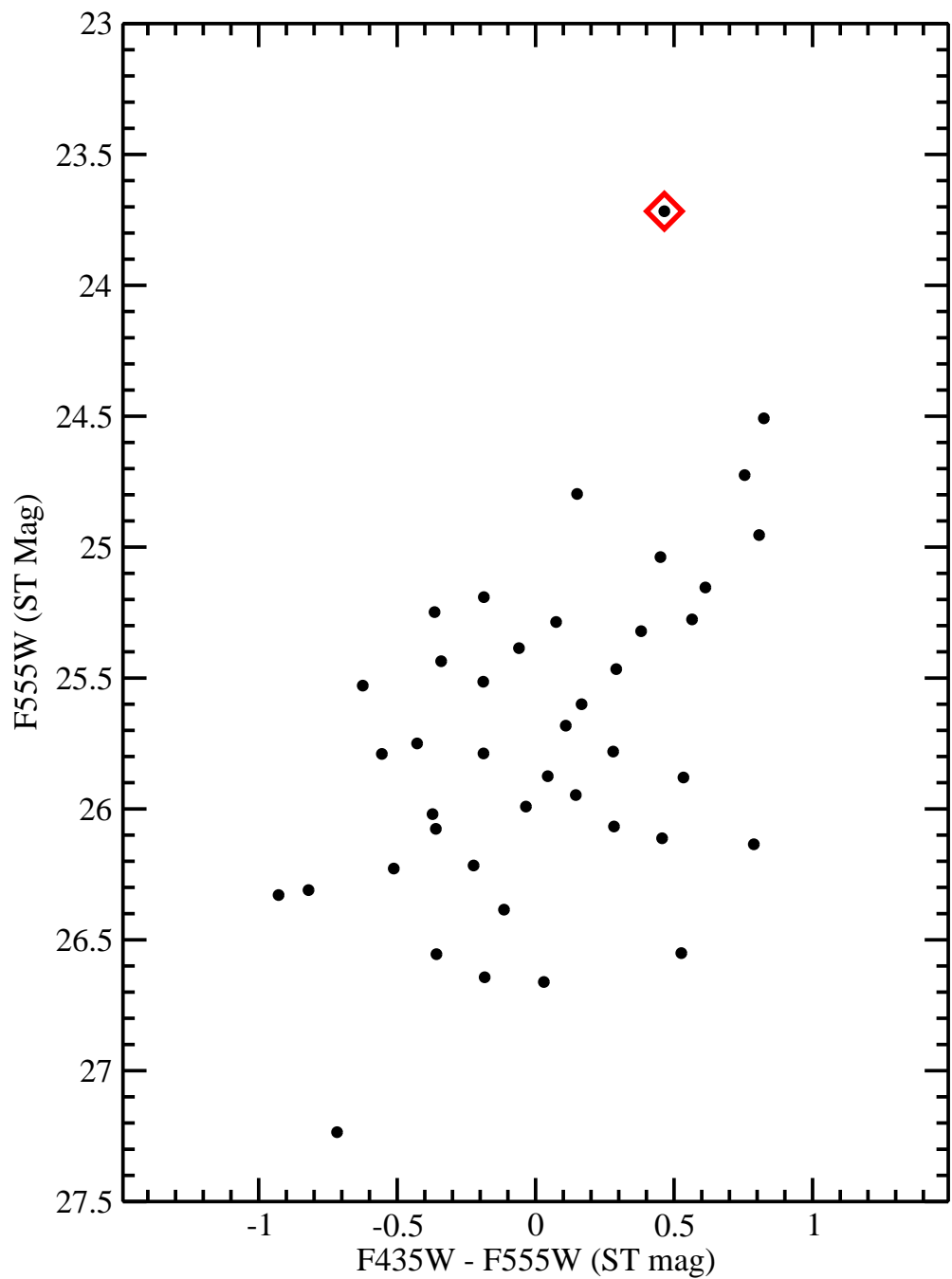}
\caption{Color Magnitude Diagram of the stars within $3 \sigma$ of the position of Bond \& Siegel (2006). The object with a diamond is circled in Figure 3; it is the only object visible in F300W within $3 \sigma$ of the position of Bond \& Siegel (2006).}
\end{figure}

\begin{figure}[t]
\figurenum{5}
\begin{center}
{\includegraphics[scale=.5]{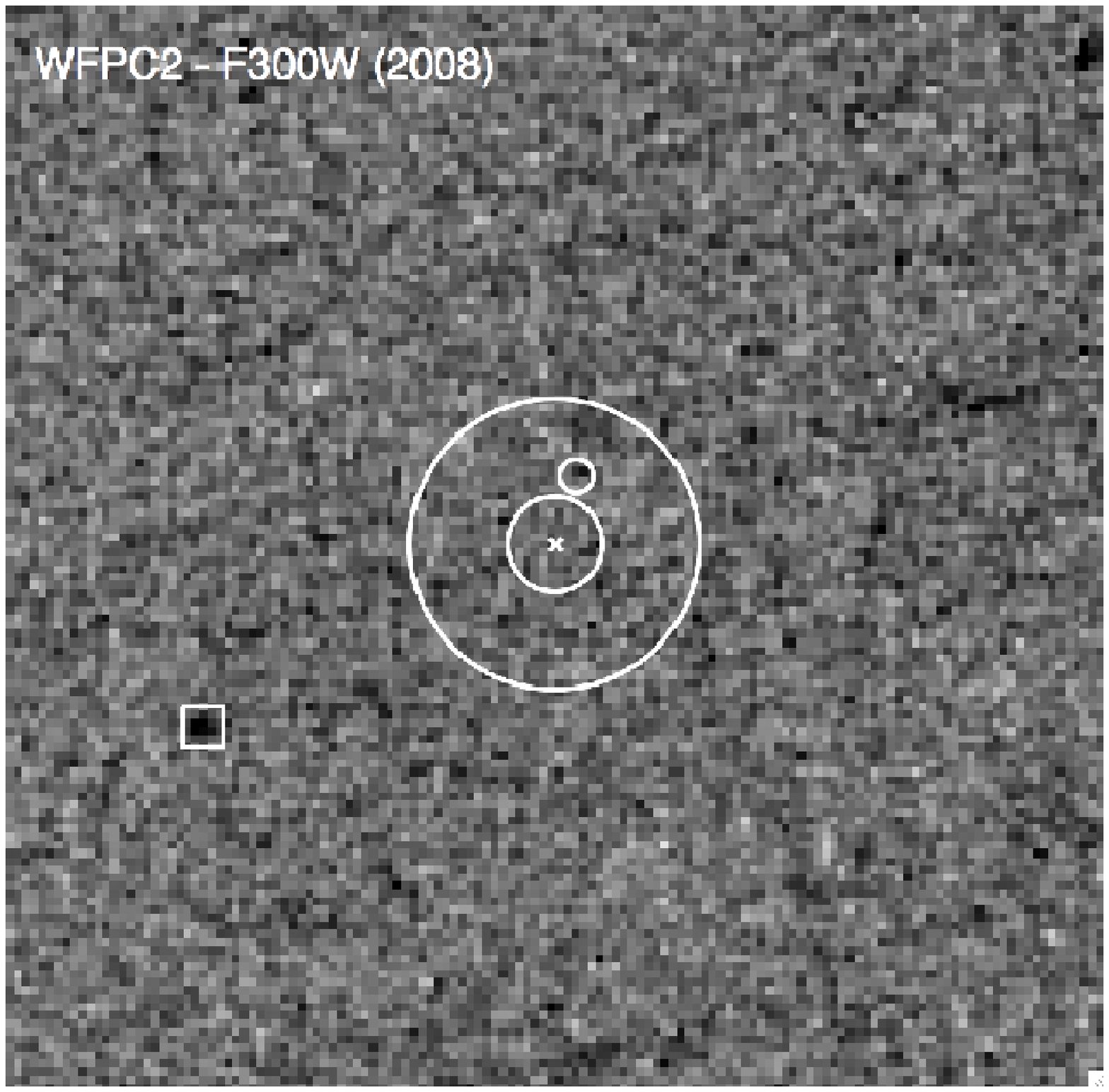}
 \includegraphics[scale=.5]{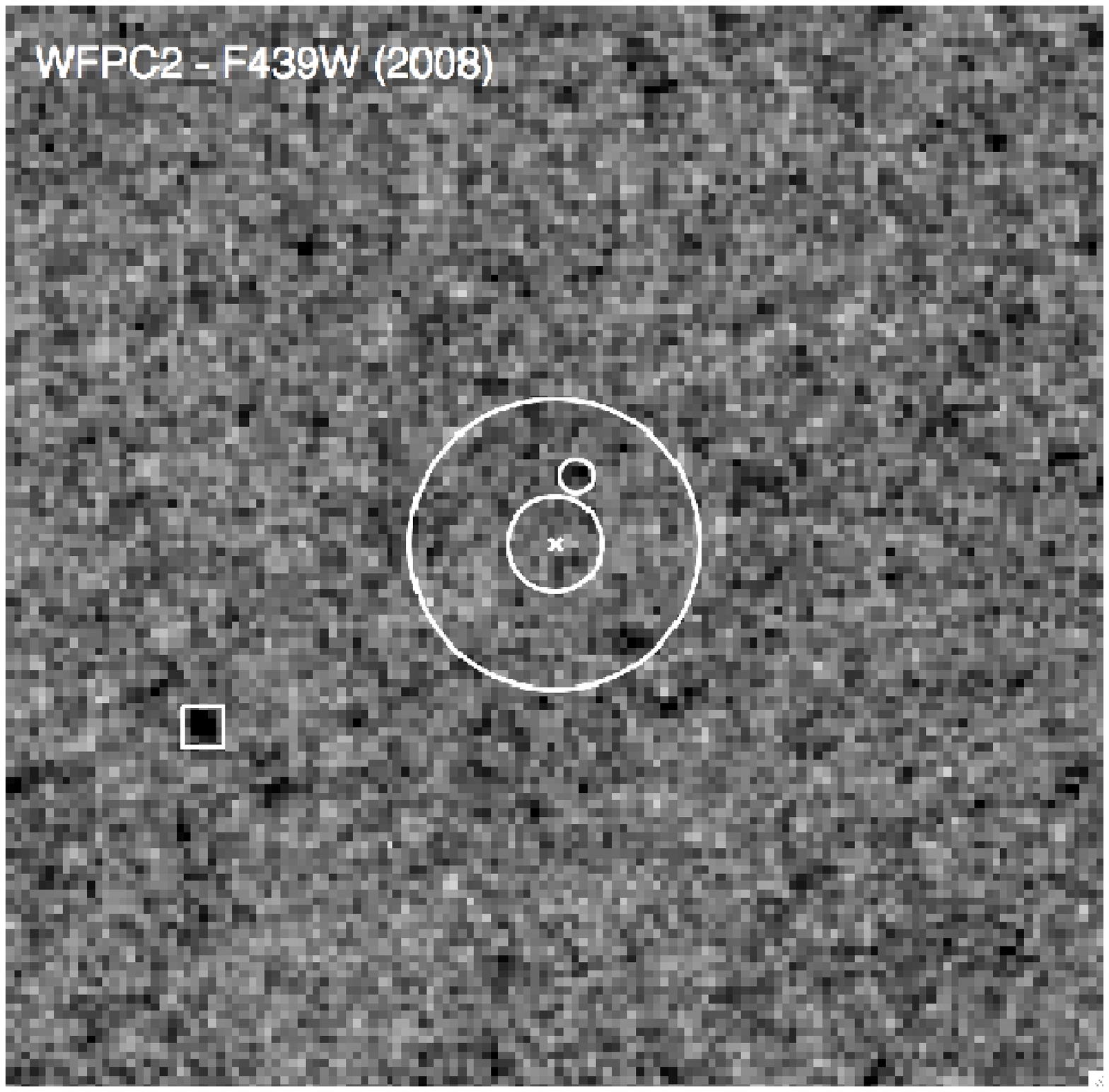}}
{\includegraphics[scale=.5]{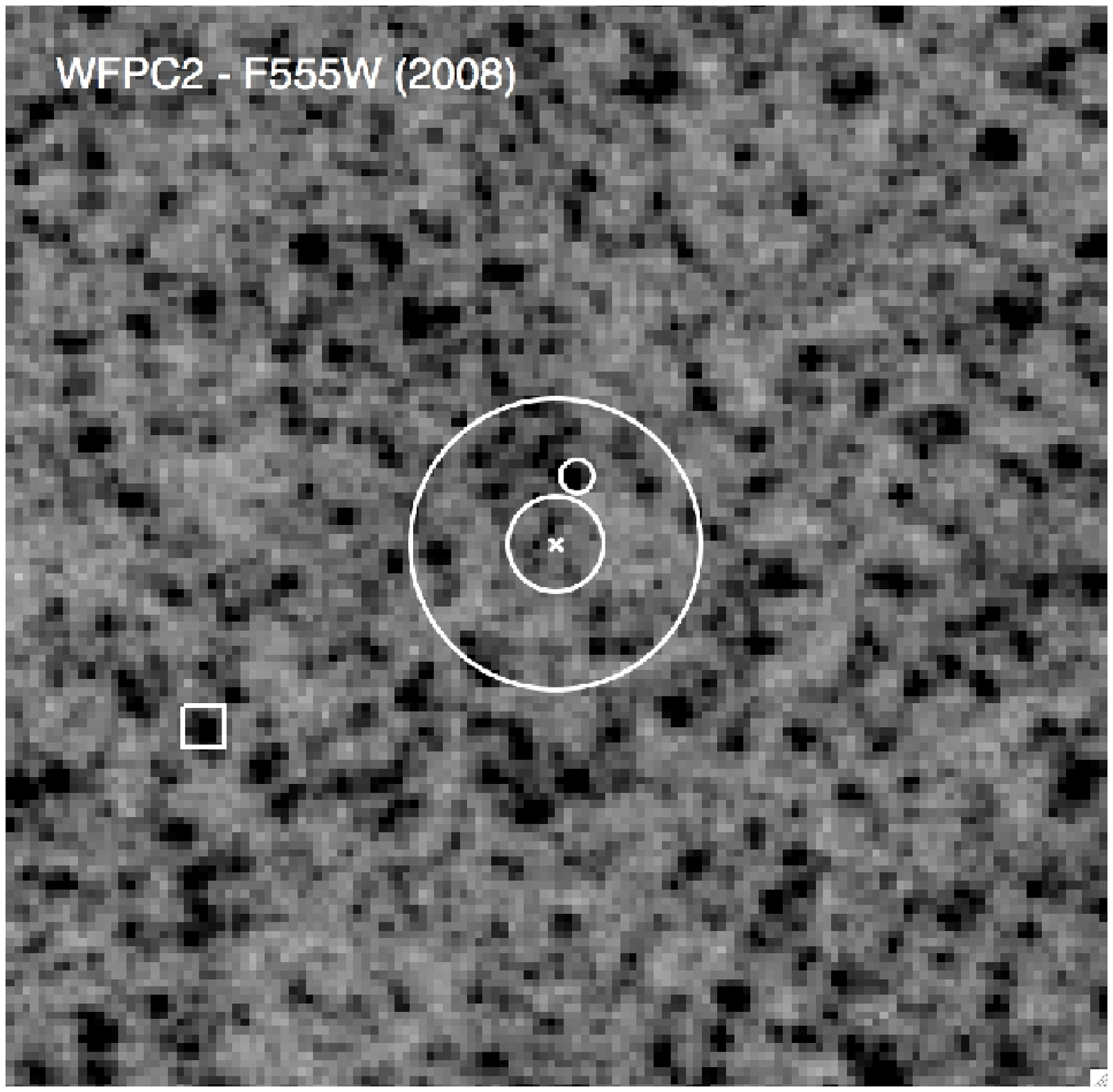}
 \includegraphics[scale=.5]{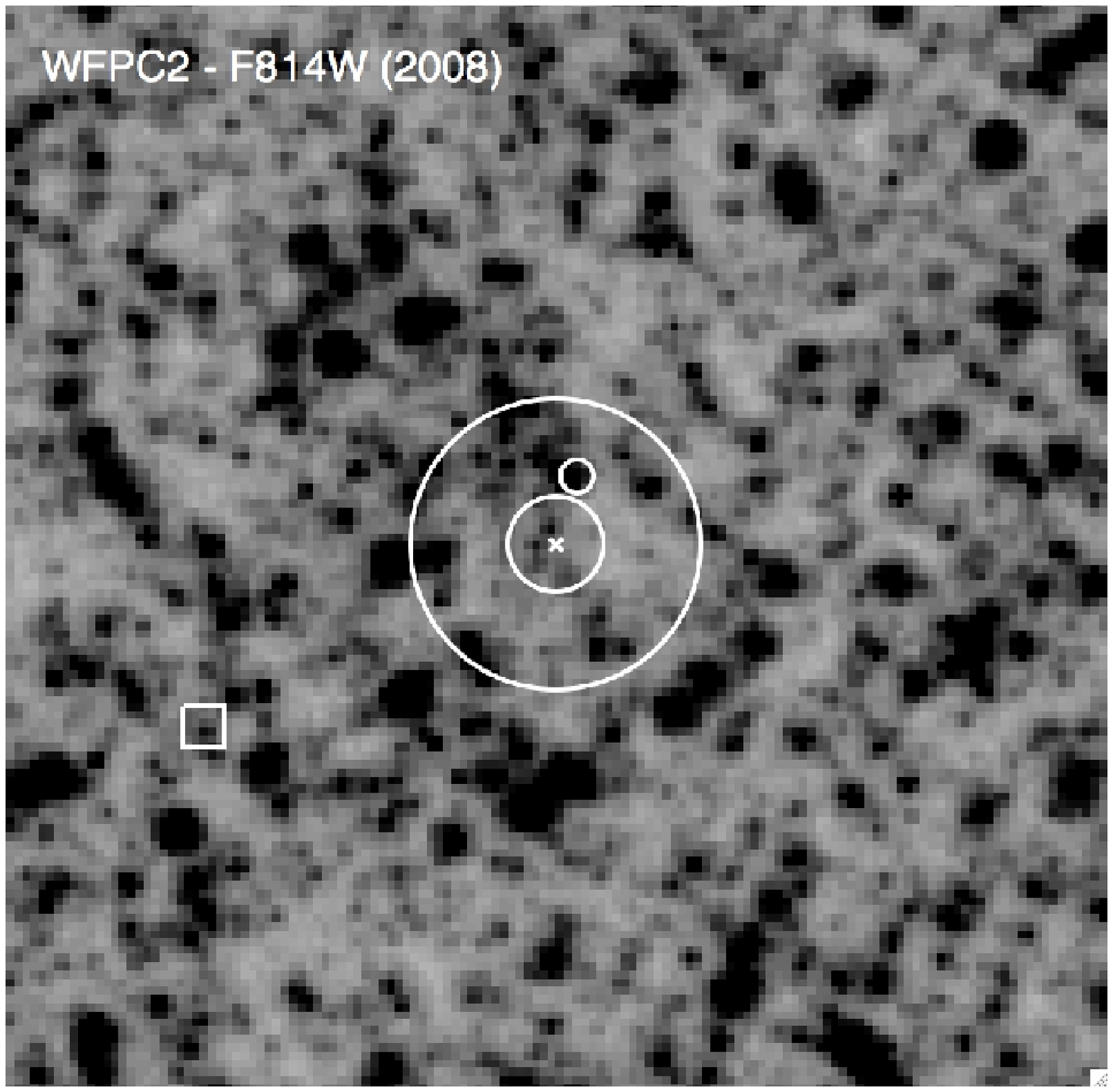}}
\end{center}
\caption{A mosaic of the F300W, F439W, F555W and F814W images taken by HST/WFPC2 in 2008.}
\end{figure}




\begin{figure}[h]
\figurenum{6}
\putplot{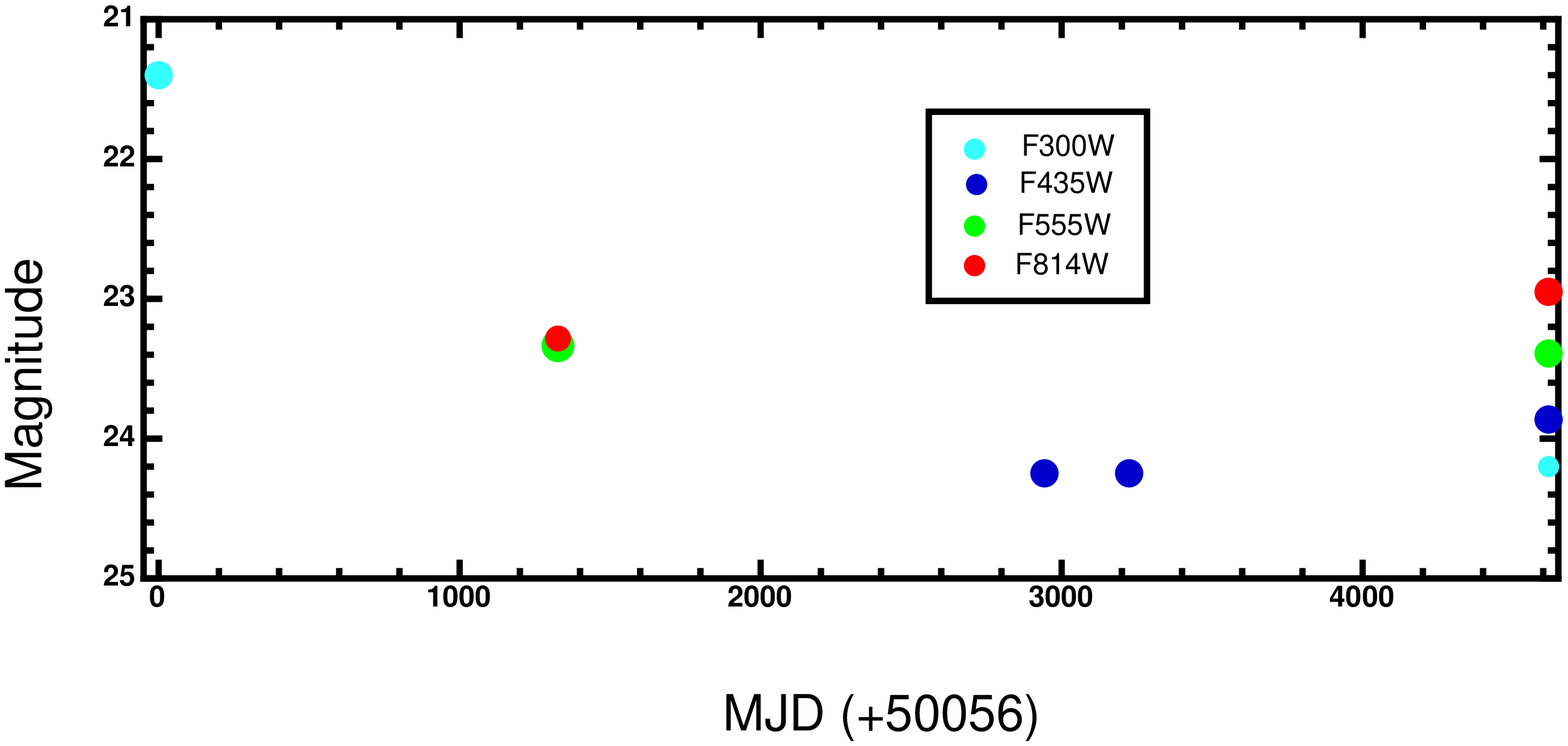}{3.0in}{0}{70}{70}{-320}{-80}
\caption{The HST light curves in all available passbands for M31-RV}
\end{figure}

\end{document}